# Accretion discs with accreting coronae in AGN – II. Nuclear wind.

Hans J. Witt,[1] Bożena Czerny[2] and Piotr T. Życki[2]
[1] *Astrophysikalisches Institut Potsdam, An der Sternwarte 16, 14482 Potsdam, Germany*, e-mail:*hwitt@aip.de*
[2] *Nicolaus Copernicus Astronomical Center, Bartycka 18, 00-716 Warsaw, Poland*, e-mail:*(bcz,ptz)@camk.edu.pl*

**ABSTRACT**
We study an accretion disc with a hot continuous corona. We assume that the corona itself accretes and therefore it is powered directly by the release of the gravitational energy and cooled by radiative interaction with the disc. We consider the vertical structure of such a corona and show that the radial infall is accompanied by strong vertical outflow. Such a model has two important consequences: (i) at a given radius the corona forms only for accretion rate larger than the limiting value and the fraction of energy dissipated in the corona decreases with increasing accretion rate, and (ii) the disc spectra are significantly softer in the optical/UV band in comparison with the predictions of standard accretion discs due to the mass loss and the decrease of internal dissipation in the disc. Both trends correspond well to the mean spectra of radio quiet AGN and observed luminosity states in galactic black hole candidates.

## 1 INTRODUCTION

Hard X-ray emission in Active Galactic Nuclei results most probably from the Compton upscatter of soft photons by a hot optically thin plasma but the place of origin of this emission is still under debate despite the amount of observational data accumulated (e.g. Bregman 1994, Czerny 1994).

A number of models were suggested. The model of a hot optically thin continuous corona above an optically thick accretion disc first discussed by Liang & Price (1977) is particularly attractive and significant progress has been recently made along this line starting mostly from the paper of Haardt & Maraschi (1991), followed by Nakamura & Osaki (1993), Kusunose & Mineshige (1994), Svensson & Zdziarski 1994, Życki, Collin-Souffrin & Czerny (1995; hereafter Paper I). It was shown that the dissipation of a significant fraction of the gravitational energy in the corona tends to stabilize the disc (Ionson & Kuperus 1984, Svensson & Zdziarski 1994). A number of papers were devoted to the stabilizing role of advection in the disc surrounded by a corona (e.g. Chen et al. 1995). The advection-dominated models were successfully applied to some galactic X-ray sources (Abramowicz et al. 1995).

Most of the previous papers were based on the assumption that the fraction of the energy generated in the corona is a free parameter of the model and does not depend on the accretion rate. This value was assumed to be constant also during time evolution computations (Abramowicz et al. 1995).

However, conclusions from the model crucially depend on this assumption (Paper I). A coupling between the fraction of the energy dissipated in the corona and the accretion rate may lead to different picture.

Observations suggest that such a coupling does exist. Direct evidence is available only for some galactic X-ray sources, namely for X-ray novae. They show rapid change of the spectrum at the end of an outburst from soft X-ray dominated spectrum to the domination by hard X-ray power law (e.g. Miyamoto et al. 1993). This suggests that when the accretion rate drops to $\sim 1\%$ of the initial value, a transition occurs from a disc-dominated solution to a corona-dominated solution (Miyamoto et al. 1993, van der Klis 1994). The initial value is probably close to the Eddington limit.

There are no such clear indications for a similar trend in Seyfert galaxies or quasars; however, the hardening of the UV spectrum of Seyfert galaxies with the increase of the luminosity may indirectly support similar correlation between the fraction of energy dissipated in the corona and the accretion rate.

Therefore a method to predict theoretically the fraction of energy liberated in the corona is of extreme importance as its predictions can be tested against observations thus allowing for deeper insight into the very phenomenon.

Such a method was outlined in the Paper I. However, that paper was based on the assumption that the corona is in a hydrostatic equilibrium which was not justified for higher accretion rates so conclusions from the model could not have been drawn.

In this paper we study a corona above an accretion disc allowing both for the vertical supersonic outflow as well as the radial advection term, with the description of the coupling between the disc and the corona as in Paper I. The description of the vertical structure of the corona including transonic vertical outflow is given in Section 2. In Section 3 we show the properties of the corona including the dependence of the calculated fraction of the energy liberated in the corona on accretion rate and viscosity, as well as we estimate the mass loss from the disc and the radial dependence of the disc accretion rate. We also compute exemplary disc/corona spectra in the optical/UV/X-ray range taking into account the mass loss from the disc. In Section 4 we discuss the importance of the radial advection. Conclusions are given in Section 5.

## 2 DYNAMICS OF THE CORONAL OUTFLOW

### 2.1 Equations of vertical motion

Following Begelman, McKee & Shields (1983) and Meyer & Meyer-Hofmeister (1994) we derive a one-dimensional model for the radial accretion and outflow process in a cylinder along the $z$-axis. We assume that the gas is isothermal and the outflow is driven by radiation pressure from the disc.

Starting with the continuity equation of mass flow in cylindrical coordinates we have

$$0 = \frac{\partial \rho v_z}{\partial z} + \frac{1}{r}\frac{\partial}{\partial r}(r\rho v_r), \qquad (1)$$

where $v_r$ and $v_z$ are the radial and vertical components of the velocity. The radial flow velocity can be approximated by

$$v_r = -\alpha \frac{c_s^2}{\Omega_K r} \quad \text{with} \quad \Omega_K = \sqrt{\frac{GM}{r^3}}. \qquad (2)$$

Here $c_s$ denotes the isothermal sound velocity ($c_s^2 = P/\rho$), $\Omega_K$ is the Keplerian angular velocity, $G$ is the gravitational constant, $M$ the mass of the black hole and $\alpha$ is the viscosity parameter.

We replace the sidewise loss term of the vertical flow by (cf. Meyer & Meyer-Hofmeister 1994)

$$\frac{1}{r}\frac{\partial}{\partial r}(r\rho v_r) \approx -\frac{2}{r}\rho v_r. \qquad (3)$$

The vertical flow is not restricted to the $z$-axis and the flow lines diverge while the geometry changes asymptotically from the cylindrical to spherical for $z > r$. This divergence is essential for the transonic outflow to develop and cannot therefore be neglected. To account for this effect, we replace the vertical mass flux with the mass flux in the diverging tube (e.g. Begelman et al. 1983)

$$\rho v_z \longrightarrow \rho v_z \left(1 + \frac{z^2}{r^2}\right). \qquad (4)$$

The introduced assumptions allow us to rewrite the continuity equation (1) in the following form

$$\frac{d(\rho v_z)}{dz} = \frac{2\rho v_r r}{r^2 + z^2} - \frac{2z\rho v_z}{r^2 + z^2}, \qquad (5)$$

where $\dot{m}(z) = \rho v_z$ now describes the vertical accretion rate per unit area. It depends only on $z$.

Second differential equation describes the motion of the gas in the corona in the vertical direction,

$$\rho v_z \frac{dv_z}{dz} = -\frac{dP}{dz} + \frac{dP_{\text{rad}}}{dz} - \frac{\rho GMz}{(r^2+z^2)^{3/2}}, \qquad (6)$$

where $P$ is the gas pressure, the last term is due to the gravity of the central object (we neglect any self-gravity effects in the disc) and the radiation pressure gradient in the corona is given by

$$\frac{dP_{\text{rad}}}{dz} = \frac{F_{\text{soft}}\kappa_{\text{es}}\rho}{c}, \qquad (7)$$

where $F_{\text{soft}}$ is the total flux emitted by the disc resulting from internal viscous generation and thermalisation of the external irradiation, and $\kappa_{\text{es}} = 0.34\,\text{cm}^2/\text{g}$ is the electron scattering opacity.

Replacing now all terms $\rho v_z$ by $\dot{m}_z$ and using the relation

$$\rho \frac{dv_z}{dz} = \frac{d\dot{m}_z}{dz} - \frac{\dot{m}_z}{P}\frac{dP}{dz}, \qquad (8)$$

we obtain two differential equation for $\dot{m}_z(z)$ and $P(z)$

$$\frac{d\dot{m}_z}{dz} = \frac{2\rho v_r r}{r^2+z^2} - \frac{2z\dot{m}_z}{r^2+z^2}, \qquad (9)$$

$$\frac{dP}{dz}\left(P^2 - \dot{m}_z^2 \frac{kT}{m_{\text{H}}}\right) = -\dot{m}_z \frac{d\dot{m}_z}{dz} P \frac{kT}{m_{\text{H}}} \\ + \frac{F_{\text{soft}}\kappa_{\text{es}}}{c}\frac{P^3 m_{\text{H}}}{kT} - \frac{P^3 m_{\text{H}} GMz}{kT(r^2+z^2)^{3/2}}, \qquad (10)$$

where $T$ is the corona temperature (equal to the ion temperature $T_i$ in the case of two-temperature plasma; see Section 2.3). These equations describe the equilibrium between accretion and outflow due to radiation pressure coming from inverse Compton scattering.

These two differential equations can be partly solved analytically as shown in the Appendix A.

The values of the temperature of the corona and the soft (disc) flux has to be determined from the energy balance (see Section 2.3).

### 2.2 Basis of the corona and the sonic point

The boundary conditions for the differential equations (9) and (10) are given by the basis of the flow pattern being the surface of an accretion disc and by the condition that the outflow has to be transonic.

The height of the disc or the basis of the corona is defined by the equilibrium of the radiation pressure and the gravitational force in vertical direction.

$$\frac{F_{\text{soft}}\kappa_{\text{es}}}{c} = \frac{GMz}{(r^2+z^2)^{3/2}}. \qquad (11)$$

The equation has two solutions for $z$ if $F_{\text{soft}}$ is smaller than

$$F_{\text{soft, max}} = \frac{2}{3^{3/2}}\frac{c}{\kappa_{\text{es}}}\frac{GM}{r^2}. \qquad (12)$$

For the limiting value the two solutions merge to one, $z = r/\sqrt{2}$. If the flux $F_{\text{soft}}$ exceeds this limit a disc may not form because the system is not in hydrostatic equilibrium. This limit corresponds to the Eddington luminosity but the numerical coefficient is slightly different due to the flat (and not spherical) geometry.

The physically relevant solution for the height of the disc, $z_0$, is taken from the range $0 \leq z_0 \leq r/\sqrt{2}$. The other solution is dynamically unstable.

The condition of the regular behaviour of the outflow at infinity is replaced, as usual, by the condition that the proper solution of the two differential equations (9) and (10) has to go through a sonic point (cf. also Bondi 1952). A sonic point is a removable singularity where both the nominator and the denominator vanish simultaneously. We therefore obtain the condition $P_{\text{sonic}}^2 = \dot{m}_{\text{sonic}}^2 kT/m_{\text{H}}$. Inserting this condition in the nominator of equation (10) yields

$$P\left[\frac{2z}{(r^2+z^2)} - \sqrt{\frac{kT}{m_{\text{H}}}}\frac{2v_r r}{(r^2+z^2)} + \frac{F_{\text{soft}}\kappa_{\text{es}} m_{\text{H}}}{ckT} \right.$$
$$\left. - \frac{GMzm_{\text{H}}}{kT(r^2+z^2)^{3/2}}\right] = 0. \tag{13}$$

This means that we can determine the position of the sonic point without knowing the explicit solution of $P(z)$, which must go through the sonic point. The sonic point $z_{\text{sonic}}$ depends only on some external quantities like the corona temperature $T$, viscosity parameter $\alpha$, radial coordinate $r$, etc.

Mathematically speaking equation (13) represents only an extension of equation (11) in which two positive terms are added (consider that $v_r < 0$ by definition). Equation (13) has not more than two solutions, like equation (11). Since the presence of the two terms on the left side is equivalent to an increase of $F_{\text{soft}}$ in equation (11), it follows that $z_{\text{sonic}} > z_0$. In addition we can notice that the sonic point must disappear if the temperature becomes too large.

### 2.3 Energy budget

Since we assume that the corona is isothermal (i.e. there is no temperature gradient in the vertical direction) our description of the energy generation and cooling basically (but not fully) follows Paper I.

We assume that both the corona and the disc dissipate the potential energy of the accreting gas. We therefore adopt the $\alpha$-viscosity description for both layers. We assume a two-temperature plasma of ions and electrons. In a steady state the heating and cooling rates must be equal. The dissipation energy of the ions is transported via Coulomb interaction to the electrons. The heating of ions arises from the viscous dissipation of the gravitational energy, while the cooling is conducted by Coulomb electron-ion exchange. Electrons heated by the last process cool by Compton scattering off soft photons from the disc. The soft photon flux comes partly from the energy generated in the disc and partly from the fraction of hard X-ray photons downscattered by the corona, as in the basic paper of Haardt & Maraschi (1991).

The basis of the corona is determined by the condition that the ionization parameter, $\Xi$, (Krolik, McKee & Tarter 1981) marginally allows the medium to reach the inverse Compton temperature (i.e. the atomic processes are negligible).

These conditions can be formulated in such a way that the structure and properties of the cool disc need not to be known explicitly.

The total generated flux of the model is given by the global parameters the black hole mass $M$, the global accretion rate $\dot{M}$, the viscosity parameter of the corona $\alpha$ and the radial coordinate $r$ where the one-zone model of the disc is considered.

The total flux is then given by

$$F_{\text{tot}} = F_{\text{d}} + F_{\text{c}} = \frac{3GM\dot{M}}{8\pi r^3}f(r), \tag{14}$$

where $f(r)$ represents the boundary condition at the marginally stable orbit

$$f(r) = 1 - (3R_{\text{Schw}}/r)^{1/2} \tag{14a}$$

in the Newtonian approximation.

For the flux generated in the disc we simply set

$$F_{\text{d}} = \xi F_{\text{tot}} = \xi(F_{\text{d}} + F_{\text{c}}), \tag{15}$$

where the fraction of energy dissipated in the disc, $\xi$, has to be calculated by iteration using the total flux and the flux generated in the corona by $\alpha$-viscosity. The flux generated in the corona by dissipation is given by

$$F_{\text{c}} = \frac{3}{2}\Omega_{\text{K}}\alpha \int_{z_0}^{\infty} P(z)\,dz = (1-\xi)F_{\text{tot}}. \tag{16}$$

The cooling of ions in the corona by the electron-ion Coulomb interaction is described by the following equation (Shapiro, Lightman & Eardley 1976)

$$F_{\text{c}} = \frac{3}{2}\frac{k(T_{\text{i}}-T_{\text{e}})}{m_{\text{H}}}\left[1+\left(\frac{4kT_{\text{e}}}{m_{\text{e}}c^2}\right)^{1/2}\right]\int_{z_0}^{\infty}\nu_{\text{ei}}\rho\,dz, \tag{17}$$

where $T_{\text{i}}$, $T_{\text{e}}$ are the ion and electron temperatures, and

$$\nu_{\text{ei}} = 2.44 \times 10^{21}\rho T_{\text{e}}^{-1.5}\ln\Lambda;\quad \text{with}\quad \ln\Lambda \approx 20 \tag{17a}$$

is the electron-ion coupling rate.

For the soft flux $F_{\text{soft}}$ we have the relation

$$F_{\text{soft}}(z_0) \equiv F_{\text{d}} + \eta F_{\text{c}}(1-a), \tag{18}$$

where $\eta$ is the fraction of the coronal flux directed towards the disc and $a$ the X-ray albedo. We assume for our model $\eta = 0.5$ and $a = 0$. The soft flux results in Compton cooling of hot electrons which is balanced by heat generation,

$$F_{\text{cool}} = F_{\text{c}} = (e^y - 1)F_{\text{soft}}, \tag{19}$$

where $y$ is the Compton parameter for optically thin medium,

$$y = \tau_{\text{es}}\frac{4kT_{\text{e}}}{m_{\text{e}}c^2}\left(1+\frac{4kT_{\text{e}}}{m_{\text{e}}c^2}\right), \tag{20}$$

and $\tau_{\text{es}}$ denotes the scattering optical thickness of the corona,

$$\tau_{\text{es}} = \int_{z_0}^{\infty} \kappa_{\text{es}}\rho\,dz. \tag{21}$$

Finally we require that the value of the ionization parameter $\Xi$ at the basis of the corona corresponds to marginal thermal stability:

$$\Xi = \Xi_0\left(\frac{T_{\text{e}}}{10^8\,\text{K}}\right)^{-3/2} = \frac{F_{\text{c}}}{2cP_0}. \tag{22}$$

Here $P_0 \equiv P(z_0)$ denotes the pressure at the basis of the corona and $\Xi_0 = 0.65$ is the ionization parameter for inverse Compton temperature $10^8$ K. We take into account that the critical value of that parameter depends on the inverse Compton temperature (Begelman et al. 1983).

Therefore in our present approach we made two important changes with respect to the thermal conditions formulated in Paper I:

(i) we define $\Xi$ using corona gas pressure instead of the disc gas pressure
(ii) we scale $\Xi$ with the electron temperature in the corona.

Both changes lead to essential changes of the character of the model.

Equations (9) and (10) with proper boundary conditions, supplemented by equations (14)–(22), allow determination of the structure and the dynamics of the hot phase including the fraction of energy liberated in the corona. The solution at a given radius is parameterized by the mass of the central object, the accretion rate and the viscosity parameter $\alpha$.

### 2.4 Cold disc structure and the stability of corona/disc solution

Although the solution is fully determined by the equations for the corona we may be also interested in calculating the properties of the cold layer in equilibrium with such a corona. The structure of the disc can be uniquely calculated by integrating equations of the disc vertical structure with the surface boundary conditions imposed by the corona solution (Czerny, Życki & Collin-Souffrin 1995). Simpler estimates based on the vertically averaged disc structure can be obtained but they are less reliable as the departure of opacities from pure electron scattering and the pressure profile influence the results.

Computations of the disc structure allow to calculate the total (i.e. disc plus corona) surface density as a function of the accretion rate for our model. This function is closely related to thermal stability of disc/corona solutions (Bath & Pringle 1982).

A coronal solution (i.e. the values of $\xi$ and $P_0$) gives the boundary conditions necessary to solve the differential equations of the disc vertical structure. We do the computations (see Section 3.5) assuming the $\alpha P_{\rm tot}$ description of viscosity. In principle, the viscosity coefficient in the disc may be different from the viscosity coefficient in the corona as the thermal conditions are actually very different in these two parts of accretion flow. However, as two values of $\alpha$ would only lead to quantitative, and not qualitative, differences from a single value model we assume the same value for the two coefficients in the present computations.

## 3  RESULTS FOR THE CORONA WITHOUT RADIAL ADVECTION

### 3.1  Topology of solutions; transonic curves

We first discuss the topology of solutions of the two differential equations (9) – (10) and the conditions of the existence of transonic curves without referring to algebraic or integral equations describing the thermal balance.

**Figure 1.** Two examples of topology of transonic solutions. The parameters are indicated in each figure. In the top panel (a) the correct transonic solution marked by a thick curve determines uniquely the initial value of $P_0$ at the basis of the corona. In the bottom panel (b) a case is shown where the transonic solution is *not* connected to the basis of the corona.

Detailed study of the equation (13) determining the position of the sonic point shows that it may have two solutions. The two solutions merge with increasing coronal temperature (the ion temperature, $T_{\rm i}$, in our two-temperature approximation), or the soft flux from the disc $F_{\rm soft}$, and then the solution disappears.

For $T \to 0$ the solutions of equation (13) converge towards the two solutions of equation (11), $z_0$ and $z_1 \approx (r/z_0 - 3/2)^{1/2}$. The solutions of equation (13) are confined between $z_0$ and $z_1$. Therefore, if the soft flux is increased, the sonic point disappears before the system reaches the Eddington limit as defined for spherical symmetry. The merging point yields an upper limit of the temperature in the corona which cannot be exceeded,

$$T_{max} = \frac{m_H c^2}{4k} \frac{R_{Schw}}{r} \left[1 - \frac{3}{2}\left(\frac{z_0}{r}\right)^{2/3}\right]$$
$$\leq 2.73 \times 10^{11} \left(\frac{10\, R_{Schw}}{r}\right) [K]. \qquad (23)$$

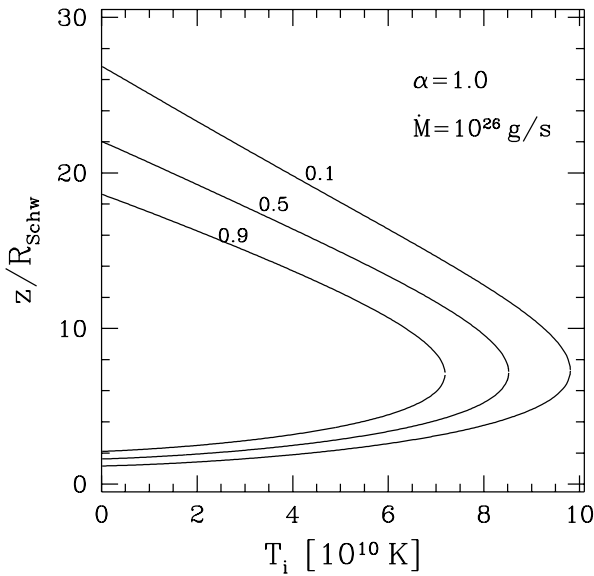
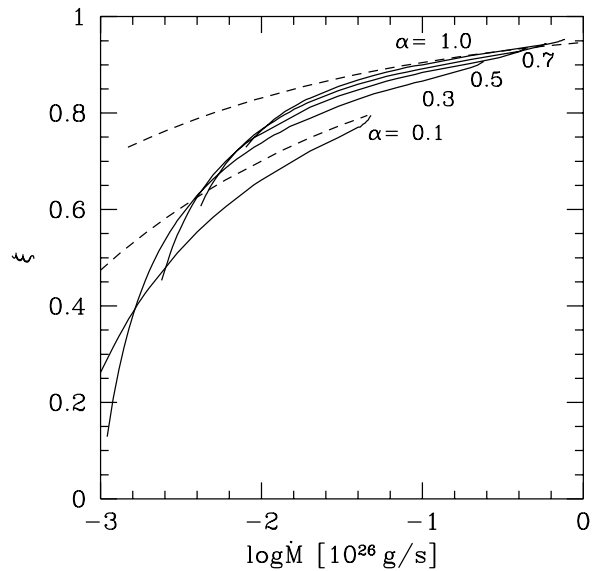

**Figure 2.** The positions of the two sonic points (see equation 13) as functions of the corona temperature $T_i$ for three values of $\xi$ (or $F_{soft}$; see equations 14, 15 and 18).

**Figure 3.** Fraction of the energy liberated in the disc as a function of accretion rate; the corona dissipates a fraction $1-\xi$. Lower limit for accretion rate corresponds to $\xi = 0$ whilst the upper limit for accretion rate and $\xi$ is due to the disappearance of transonic solutions. The curves are labeled by the value of the viscosity parameter $\alpha$. The dashed curves show the analytical approximation given by equation (25) for $\alpha = 0.1$ and 1. Other parameters are: $M = 10^8$ M$_\odot$, $r = 10\, R_{Schw}$.

The formula is derived for $\alpha = 0$ and small $z_0$. The maximum possible temperature in the corona is smaller for higher viscosity parameter $\alpha$.

The solution close to the disc surface is positioned at $z_0 < z \leq z_{merge}$, where the merging point cannot exceed the limit $z_{merge} \leq 0.835r$. The corresponding singularity is of the *focus* type. There is no transonic solution passing through such a singularity. Physically it is related to the fact that at this position the gravitational potential still increases outwards so there is no acceleration of the outflow. The other solution is located at $z_{merge} \leq z < z_1$ where the gravitational potential decreases outwards so the velocity becomes transonic. This type of singularity is called a *saddle* and it allows to cross from subsonic to supersonic motion. The coexistence of the two singular points is a typical property of a flow with a potential barrier, e.g. in the case of disc accretion in the vicinity of the marginally stable orbit (Muchotrzeb 1983).

Two examples of the topology of the solutions of equations (9)–(10) are shown in Fig. 1. These figures were derived by varying the integrating constant (see for details Appendix B), i.e. a number of different initial values of the pressure was chosen. In the top panel (a) we show an example of the existence of the disc/corona solution. There is a curve which starts at the disc surface at high pressure and low velocity values, and passes through the sonic point at $z \approx 17.7$ achieving, with increasing $z$, low pressure and high velocity values. In the bottom panel (b) we show a situation where the solution which is going through the sonic point is not connected to the basis of the corona. For such parameter space there is no uniquely physical solution. For the following discussion it is important to understand that the family of physical solutions is limited by the fact that either the two sonic points converge or the transonic curve cannot be accessed from the disc surface.

The convergence of the two sonic points is illustrated in Fig. 2.

### 3.2 Fraction of energy liberated in the corona

When the results from the previous section are supplemented by the equations of energy budget (see Section 3.3), the soft flux or, equivalently, $\xi$ are no longer free parameters.

In Figure 3 we show the dependence of $\xi$ on the accretion rate. The computations were done for central black hole mass $10^8$ M$_\odot$, at a radius of 10 $R_{Schw}$. The curves correspond to different values of the viscosity parameter $\alpha$ ranging from 0.1 to 1.

The most striking differences between the relation $\xi$ vs. $\dot m$ obtained here and in Paper I are: (i) the existence of one solution here instead of two in Paper I and (ii) the basic solution of Paper I predicted the increase of the relative importance of the corona with accretion rate whilst our solution predicts the opposite trend, (iii) the solutions exist for much broader, albeit still limited, range of accretion rates.

These differences result mostly from the two changes introduced to the consideration of the energy budget and not from the inclusion of the dynamical terms into the corona equations.

First, we replaced the gas pressure at the disc surface with the corona pressure at the basis in relation (22). Disc-dominated models obtained by Paper I had $\xi \sim 1$ and the disc surface pressure was strongly dominated by the radiation pressure so the total pressure equilibrium did not result in gas pressure equilibrium at the disc/corona boundary. Our new definition of the disc/boundary layer seem to be more correct as we use the location of the upper bend on the $T$ vs. $\Xi$ curve to define the transition (Krolik et al. 1981; see also Paper I). However, the problem indicates the necessity of the more detailed computations of the transition zone.

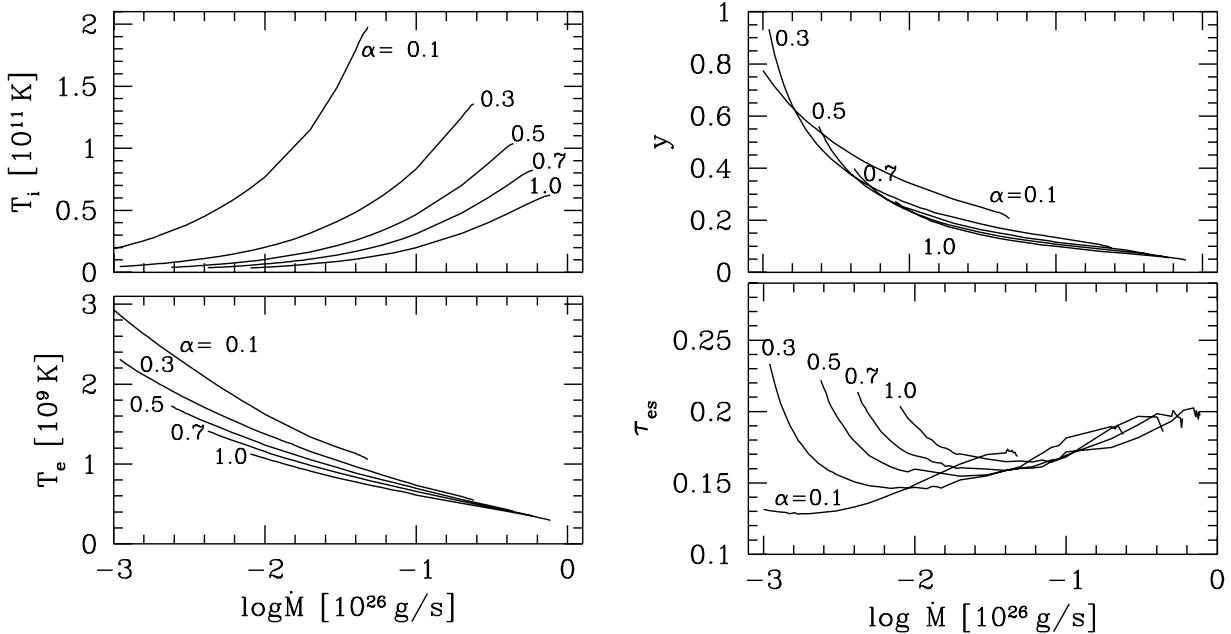

**Figure 4.** The dependencies of corona parameters: the ion temperature, $T_i$, the electron temperature, $T_e$, the electron scattering optical depth, $\tau_{es}$, and the Compton parameter, $y$, on the accretion rate, $\dot{M}$, for a number of values of the viscosity parameter $\alpha$. Model parameters: $M = 10^8\,M_\odot$, $r = 10\,R_{\rm Schw}$.

Using our present definition of $\Xi$ with some other simplifications ($T_i \gg T_e$, $H_c \ll r$, $4kT_e/m_e c^2 \ll 1$), we obtain an approximate relation for $\Xi = \Xi_0$,

$$\dot{M} = A(1-\xi)^{-1} \left(\ln \frac{3-\xi}{1+\xi}\right)^{3/5}, \qquad (24)$$

where

$$A = 2.5 \times 10^{22}\,\Xi_0^{12/5}\,\alpha^{-1}\,\Omega_K^{-1}\,f(r)^{-1}\;[\text{g/s}].$$

This equation shows that the fraction of the energy generated in the corona increases with the accretion rate. However, the assumption $\Xi = \text{const}$ leads to relatively narrow range in $\dot{m}$ with interesting solutions. These solutions exist only for relatively low values of $\Xi_0$ ($\sim 0.01$); otherwise the accretion rate exceeds the Eddington rate.

Much broader range of discs with relatively strong coronae exist if the dependence of $\Xi$ on the coronal temperature is included, as we do in the present model (see equation 22). An approximate solution can be derived also in this case:

$$\dot{M} = A(1-\xi)^{-1} \left(\ln \frac{3-\xi}{1+\xi}\right)^{-3}, \qquad (25)$$

where

$$A = 8.3 \times 10^{15}\,\alpha^{-1}\,\Omega_K^{-1}\,f(r)^{-1}\;[\text{g/s}]$$

Although these solutions approximate relatively well the exact solutions for intermediate values of $\xi$, they misrepresent them for both very high and very low accretion rates. For very low accretions the correction terms due to the high value of the electron temperature are important, since the electron temperature varies as $T_e \sim \dot{m}^{-1/2}$ (see Appendix C).

In the case of high accretion rates the dynamical terms become important. However, their inclusion in the corona structure does not extend the existence of solutions into much higher accretion rates than that allowed by the hydrostatic equilibrium models (see Paper I). The physical mechanism limiting the accretion rate is however different from the one discussed by Paper I in the case of hydrostatic equilibrium solutions. Now the two sonic points converge (see Section 3.1) with increasing $\dot{m}$ because both $T_i$ and $F_{\rm soft}$ increase with $\dot{m}$. There are no stationary transonic solutions above the merging point i.e. for higher $\dot{m}$, and this conclusion is not changed if the advection term is taken into account (see Section 4) The limit depends on the viscosity parameter $\alpha$, as can be seen from Fig. 3.

With decreasing accretion rate the relative importance of the corona increases. The limit $\xi = 0$, i.e. all energy generated in the corona (in this case the disc radiates only at the expense of irradiation by corona), is reached for finite value of the accretion rate. For radial distance $10\,R_{\rm Schw}$ and $M = 10^8\,M_\odot$ this limit is $\sim 8 \times 10^{22}$ g/s. For lower accretion rates there are no solutions for a disc with a corona.

As the fraction of energy liberated in the corona depends explicitly also on the current disc radius, this fraction is a function of a radial distance for a given accretion rate. We show the corresponding plot (Fig. 6) under the assumption that the accretion rate is constant, i.e. the vertical losses are relatively unimportant. The corona covers only a limited fraction of a disc ($r < r_{\rm max}$) which increases with the accretion rate. It develops mostly near $r_{\rm max}$ and its relative importance decreases inwards.

### 3.3 Properties of the corona

In Fig. 4 we illustrate the properties of the corona. The computations are made at a radius 10 $R_{\rm Schw}$, assuming $M = 10^8 \, {\rm M}_\odot$.

The optical thickness of the corona is moderate ($\sim 0.1 - 0.2$) so the assumption of optically thin medium is well satisfied. The electron temperature is never very high so there is no significant pair creation process. The ion temperature is mostly higher than the electron temperature although the difference between them is lower for low accretion rate. As a result, the value of the Compton parameter $y$ is never very high; it is close to 1 only for accretion rate close to the minimum value and is decreases from 0.6 to 0.05 for accretion rates increasing from $10^{23}$ to almost $10^{26}$ g/s.

Approximate analytical solutions illustrating the dependence of the coronal properties on the disc radius are given in Appendix C. Those formulae are not very accurate because at higher accretion rate the departure from hydrostatic equilibrium becomes essential and at low accretion rate the nonlinear terms in equations(17) and (20) are important. However, as the dynamical effects are not important at low accretion rates approximate numerical results can be found without the necessity to solve numerically the transonic outflow problem (see Section 3.4 and Appendix D).

### 3.4 Vertical vs. radial flow and global solutions

The relative importance of the vertical outflow can be conveniently measured by the asymptotic value, $\dot{m}_z^\infty$, of the expression $\dot{m}_z(1+z^2/r^2)$ normalized to a logarithmic derivative of the mass flux at a given radius, i.e.

$$\frac{d\ln \dot{M}}{d\ln r} = \frac{4\pi r^2 \dot{m}_z^\infty}{\dot{M}}. \qquad (26)$$

Numerically computed values at $10 R_{\rm Schw}$ as a function of accretion rate are shown in Figure 5. We see that the mass loss strongly depends on the accretion rate – it increases significantly with the luminosity assuming constant viscosity parameter $\alpha$. What is important, it also strongly depends on the value of the viscosity parameter $\alpha$. For $\alpha$ close to 1 mass loss is never of any importance whilst for $\alpha$ of order of 0.1 it may cause the outflow of the major (dominating) fraction of the mass flux. As the value of viscosity is not known the detailed predictions of the mass loss are impossible along this line. We can only show a few examples of the possible consequences of the outflow and, if the overall detailed predictions of the model are determined and confronted with the data for a range of values of $\alpha$, the viscosity parameter can be constrained independent from other estimates (e.g. Siemiginowska & Czerny 1989).

We now consider in some detail the case of $\alpha = 0.1$. The mass loss from the corona is clearly important and the global solution for the disc cannot be obtained assuming $\dot{M} = \rm const$ in the region where a corona develops. The correct solution has to be obtained numerically by computing the mass loss

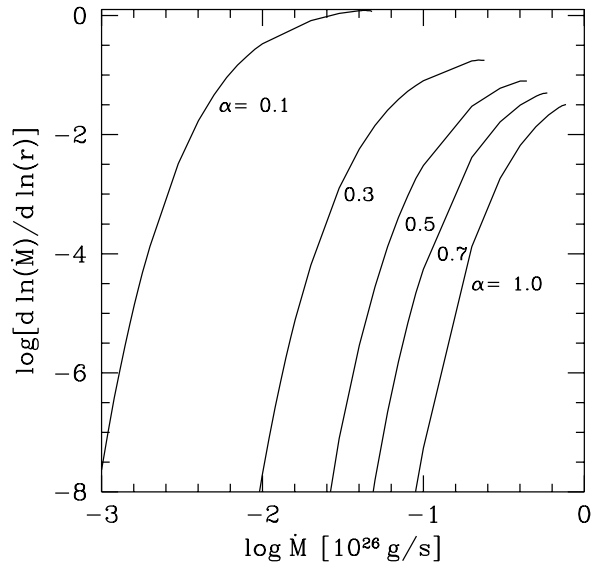

**Figure 5.** The dependence of the vertical outflow ratio as a function of accretion rate (lower axis) or the fraction of energy liberated in the corona $(1-\xi)$ (upper axis). Curves are labelled by the viscosity parameter $\alpha$. Other parameters are : $M = 10^8 \, {\rm M}_\odot$, $r = 10 \, R_{\rm Schw}$.

at every radius and then the disc/corona structure for the reduced accretion rate.

As the construction of the exact solutions at every radius and a range of accretion rates (as now the local accretion rate is not known *a priori*) is time-consuming we use for that purpose half-numerical and half-analytical approach to estimate the local mass loss rate.

We compute first the corona parameters neglecting dynamical terms (but treating carefully the equation of hydrostatic equilibrium); we also include all nonlinear terms (see equations 17 and 20) as the mass loss in very sensitive to the obtained ion temperature. The set of equations (D4)–(D7) is given in the Appendix D. These equations yields us the approximate global solutions $\xi$, $T_{\rm i}$, $T_{\rm e}$ and $P_0$ as functions of the parameters $M$, $\dot{M}$, $\alpha$ and $r$.

Using now equations (11) and (13) we can find the basis of the corona $z_0$ and the sonic point $z_{\rm sonic}$. Finally with equation (A6) we can estimate the pressure at the sonic point $P_{\rm sonic}$ which subsequently yields $\dot{m}_{\rm sonic}$.

Now we can use very good approximation,

$$\frac{d\ln \dot{M}}{d\ln r} = \lim_{z \to \infty} \frac{4\pi}{\dot{M}}(r^2 + z^2)\dot{m}_z \\ \approx \frac{4\pi}{\dot{M}}(r^2 + z_{\rm sonic}^2)\dot{m}_{z,\rm sonic}, \qquad (27)$$

which gives us the mass flux at a given radius $r$.

This local value of the mass loss as well as the determination of the local fraction of the energy liberated in the disc is incorporated into the code calculating the spectrum of a Keplerian disc radiating locally as a black body. Computations are done inwards which is equivalent to numerical integration of equation (26) starting from a given external accretion rate.

As the fraction of energy generated in the corona increases for decreasing accretion rate at a given radius but it also increases for a constant accretion rate and decreasing radius (see equation 25) we have two competing trends. The net effect is presented in Fig. 6. We see that the relative importance of corona decreases inwards although more slowly than in the case of negligible mass loss from the disc surface.

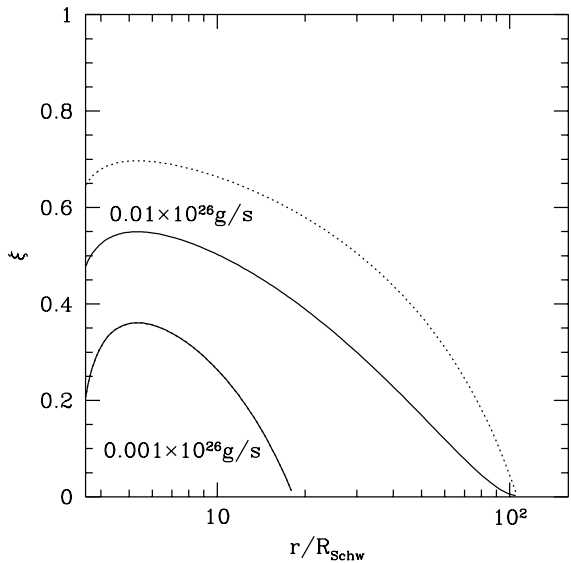
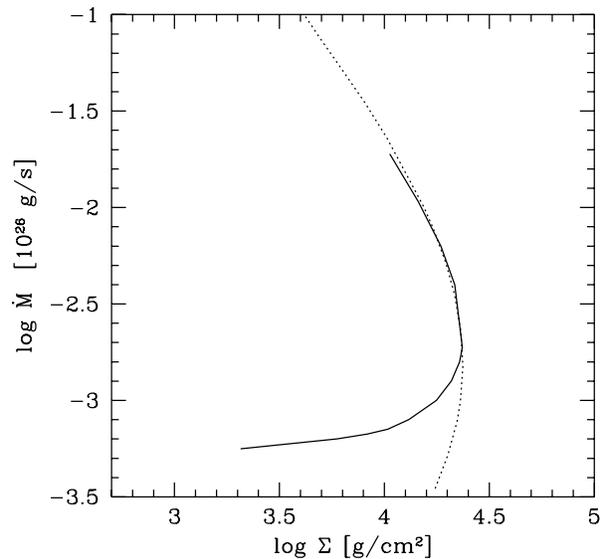

**Figure 6.** Fraction of the energy liberated in the disc, $\xi$, as a function of radius, $r$, for two values of the external accretion rate, 0.001 and 0.01 in units of $10^{26}$ g/s. The dashed curves represent the solution with vertical mass loss neglected whilst the solid curves show the solution with the effect of inward-decreasing accretion rate taken into account (the solid and dashed curves overlap for the lower value of $\dot{M}$). The viscosity parameter $\alpha$ is equal 0.1.

**Figure 7.** Relation: total surface density of the coupled disc/corona system, $\Sigma$, vs. accretion rate, $\dot{M}$. Positive slope of the curve corresponds to thermally stable solutions whilst a negative slope indicates thermal instability at a given stationary accretion rate. The viscosity parameter $\alpha = 0.1$. Other parameters are: $M = 10^8$ M$_\odot$, $r = 10\, R_{\rm Schw}$.

The presented results, however, should be treated only as an indication of a trend, and not as strictly quantitative predictions. Further work towards better approximation of this two-dimensional flow is necessary.

### 3.5 Thermal stability of the disc/corona system

The total surface density of the disc/corona system resulting from integration of the vertical structure of the cool part of the flow (see Section 2.4) is shown in Fig. 7.

The slope of the presented curve is negative for higher accretion rates. Therefore, although the presence of the corona suppresses the well known thermal instability of radiation pressure dominated discs, due to the irradiation (e.g. Czerny, Czerny & Grindlay 1986) and the decrease of energy liberation (e.g. Svensson & Zdziarski 1994), it does not remove the instability for higher accretion rates in our models.

The problem is not a particular property of our model, but a general difficulty to accommodate two opposite requirements into the model. As the data clearly suggest (see Section 3.6.1) the UV bump should be more profound at higher accretion rates, i.e. the corona should be weaker at higher accretion rates. But weaker corona cannot stabilize the disc. Therefore, at least within the frame of the $\alpha P_{\rm tot}$ description of the disc interior, either we can, at least in principle, construct models which produce stable discs with stronger corona for higher accretion rates or unstable discs with weaker corona for higher accretion rates. The first type of models is consistent with observed spectral trends but opens the question of the development of thermal instabilities within the disc, particularly the problem of the final state. The second type of models give stable models but with spectral trends inconsistent with the data. This kind of difficulties led to suggestions of the viscosity parameterization based on the $\alpha P_{\rm gas}$ model with variable (basically bimodal) $\alpha$ (Chen & Taam 1994) leading to two stable branches and an unstable branch at intermediate accretion rates.

### 3.6 Relevance to observations

#### 3.6.1 Big bump versus hard X-ray luminosity

The observed multi-wavelength variability in many Seyfert galaxies and radio quiet quasars is characterized by hardening of the UV spectrum with an increase of brightness (e.g. Ulrich 1989) and mostly constant hard X-ray slope, particularly if corrections for the warm absorber and/or reflection from cold gas is taken into account (see Mushotzky, Done & Pounds 1993). Extensive monitoring campaigns for a few sources suggest that two distinctive kinds of variability are observed. Short term variability (days - months) is generally characterized by UV flux being well correlated with the $2 - 10$ keV flux (see Clavel et al. 1992 for NGC 5548. However, sudden increase of the UV or soft X-ray flux is sometimes observed without the corresponding increase in hard X-rays (for NGC 5548 see Clavel et al. 1992; Done et al. 1995).

The first kind of variability is most probably due to some instabilities in the corona dominated flow assuming that the UV radiation originates mostly from reprocessing of X-rays (e.g. Rokaki, Collin-Souffrin & Magnan 1993). The second kind of behaviour reflects rather sudden increase of accretion rate in the innermost part of accretion disc; the timescale of 1 week for NGC 5548 is consistent with viscous timescale of the innermost parts of the disc for viscosity parameter $\alpha \sim 0.1$, assuming the disc thickness to the radial distance ratio about 0.005 (Siemiginowska, Czerny & Kostyunin 1995), the value of the central mass $3.7 \times 10^7 M_\odot$ (Clavel et al. 1992) and using the thermal time scale as derived by Siemiginowska & Czerny (1989).

Our model correctly reproduces the trends accompanying the increase of accretion rate. Particularly if the accretion rate is close to the minimum value, then minor changes in accretion rate can drive major spectral changes.

### 3.6.2 UV spectrum

The existence of the corona modifies the shape of the optically thick emission through the decrease of the energy dissipation in the disc as well as through Compton scattering of the outgoing photons. In this section we concentrate on the first effect.

According to our model, the optically thick emission (i.e. soft flux; see Section 2.3) is determined by the local value of $\xi$ and $\dot{M}$. These quantities are determined by equations (27) and (C1) and the mass loss from the disc is calculated as described in Section 3.4. Disc spectrum calculated under the assumption that the local emission is well approximated by black body is shown in Fig. 8. More advanced disc spectra are also available in the literature (e.g. Czerny & Elvis 1987, Laor & Netzer 1989, Ross, Fabian & Mineshige 1992, Shimura & Takahara 1995) but as the accuracy of the description is still problematic due to simplifications in the description of the disc vertical structure and the radiative transfer, we decided to use the simplest model for the purpose of illustration.

The examples of the resulting spectra (without Comptonization by the corona) are shown in Fig. 8. They were calculating assuming Keplerian disc locally radiating as a black body but both the mass loss (see Section 3.4) and the decrease of the internal dissipation due to the presence of the corona (equation 18) were taken into account.

As the accretion rates drops by a significant factor on the way to the black hole the overall spectrum is flatter in the optical band than the usual spectrum of a stationary conservative Keplerian disc.

In our case the mass loss (and subsequently the energy index $\alpha_E$, $F_E \sim E^{-\alpha_E}$ in UV) is larger for larger accretion rate. However, this trend may reverse if there is any change of the viscosity parameter with the corona temperature or bolometric luminosity.

### 3.6.3 X-ray spectrum

For most values of external accretion rates we predict significant Comptonization in the corona if the mass loss from the disc is taken into account. The values of the Compton parameter $y$ are typically $0.5 - 0.7$, as can be seen from Fig. 6 ($y \sim 1 - \xi$). As our computations have still approximate character we did not compute the resulting spectra taking into account the radial properties of the corona. Instead, we calculated the accretion rate at 10 $R_{\rm Schw}$ appropriate for assumed external accretion rate, computed the corona parameter at this radius and Comptonized the total disc spectrum assuming a uniform corona with properties as discussed above. At the present stage we do not include the X-ray radiation created in the corona but reflected by the disc before reaching an observer.

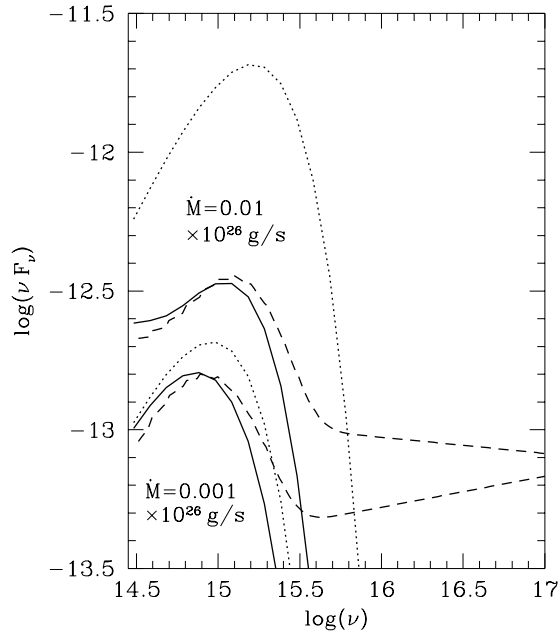

**Figure 8.** Two examples of the spectral shape of the optically thick, local black body emission of a stationary, Keplerian accretion disc with both the effect of $\xi < 1$ and decreasing (with $r$ decreasing) local accretion rate due to the outflow (*solid curves*). The external accretion rates are 0.001 and 0.01 in units of $10^{26}$ g/s. The dashed curves show effects of Comptonization of the spectra by the corresponding coronae (see text for details). For comparison, the spectrum of a conservative disc ($\dot{M}(r) = $ const) without a corona ($\xi = 1$) is shown with a dotted curve for the same two values of the accretion rate.

The result of the Comptonization is the formation of the hard X-ray power law tail with energy index 0.92 for lower value of the external accretion rate ($0.001 \times 10^{26}$ g/s) and 1.35 for the 10 times higher value. This first value corresponds well to the observed spectra of Seyfert galaxies if they corrected for the reflection component (Pounds et al. 1990; Nandra & Pounds 1994). The second value may be adequate for a few exceptional Seyfert galaxies like RE 1034+39 (Pounds et al. 1995) and some quasars.

The predicted spectra, however, depend strongly on the estimates of the mass loss from the disc surface which in turn strongly depends on the viscosity parameter (see Section 3.4). If the actual mass loss is lower the corona is weaker and the hard X-ray power law of the required slope would not form by Comptonization in a continuous corona. In that case the hard X-ray emission should originate in compact active regions developing in the corona most probably due to reconnections of the magnetic field lines, as recently discussed by Haardt, Maraschi & Ghisellini 1994 and Stern et al. 1995.

In any case the Comptonization by the corona leads to modification of the high frequency tail of the big bump leading to a power law shape instead of thermal exponential cut-off. Actually, in most sources with very strong soft

X-ray excesses the spectrum looks like an extension of the thermal emission of the big bump but the slope is a power low instead of an exponential cut-off (e.g. Czerny & Elvis 1987, Walter & Fink 1994). On the basis of EUVE and ROSAT WFC observations Marshall, Fruscione & Carone (1995) argue that the typical Seyfert spectrum in the X and UV range is given approximately by a power law with energy index $\sim 1.16 - 2.07$. Such a shape of the spectrum is easily explained by Comptonization by the plasma with a low to moderate value of the Compton parameter. Our continuous corona is therefore in a natural way identified with this medium.

Further more detailed modeling is necessary before we can distinguish between different scenarios of the hard X-ray production and the origin of soft X-ray emission for a given source. Variability studies will be particularly helpful for that purpose (e.g. Czerny, Życki & Loska 1995).

## 4 RESULTS FOR THE CORONA WITH RADIAL ADVECTION

### 4.1 Role of advection

The set of our equations describing the heating/cooling of the corona (see Section 2.3) does not take into account the radial energy transport. The most important term is the radial advection; its importance in the energy budget of the fast accretion flow was discussed in a number of papers (Muchotrzeb & Paczyński 1982; Muchotrzeb 1983, Taam & Lin 1984; Abramowicz et al. 1988; Chen & Taam 1993; Narayan & Popham 1993; Narayan & Yi 1994,; Chen 1995). In the case of coronal accretion the corresponding expression for advective heating is given by

$$4\pi r F_{\rm adv} = (1-\xi)\dot{M}T_{\rm i}\frac{dS}{dr}, \qquad (28)$$

where $(1-\xi)\dot{M}$ is the local accretion rate in the corona, $T_{\rm i}$ is the temperature (i.e. the ion temperature in the case of two temperature medium and $S$ is the mean entropy in the vertical direction. For a gas pressure dominated medium this equation can be conveniently expressed as

$$4\pi r^2 F_{\rm adv} = (1-\xi)\dot{M}c_{\rm s}^2 \delta; \qquad \delta = (2.5\frac{d\ln T_{\rm i}}{d\ln r} - \frac{d\ln P}{d\ln r}). \quad (29)$$

The sign of this term depends on the enthropy gradient in the flow. We can estimate it assuming that the advection is negligible, i.e. on the basis of solutions presented in Section 3. The value of the coefficient $\delta$ is approximately

$$\delta = -3/4, \qquad (30)$$

mostly due to the fact that the ion temperature is a decreasing function of radius and the cooler gas flows into the hotter coronal region thus acting as a cooling mechanism. We adopt the expression (29) with the value of $\delta$ given by (30) and we recalculate the structure of the corona adding the advection term to the viscous dissipation term in equation (16). Similar method was used while computing the role of advection in the disc itself (Chen 1995).

### 4.2 Properties of the corona

The inclusion of the advection term, as described in Section 4.1., does not change the results significantly. The advection is unimportant at low accretion rates. Increasing accretion rate we still first meet the topological problems with solutions – first the transonic curve does not pass through the first singular point and soon after the two sonic points converge (see Section 3.1) – and the solutions cease to exist before the advection term has a chance to dominate the thermal balance of the flow.

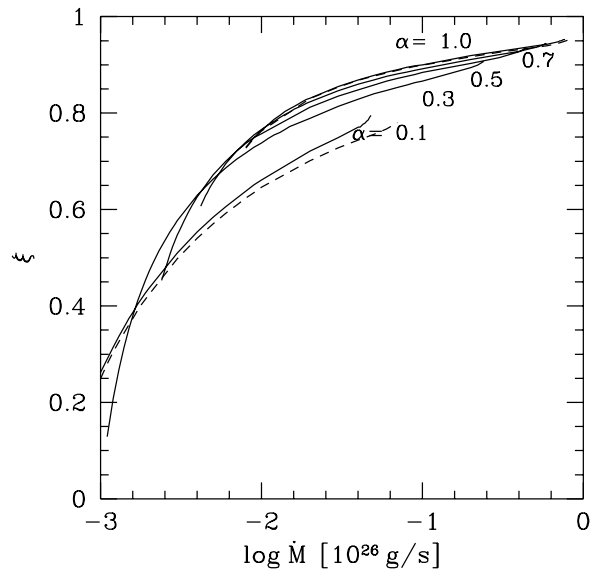

**Figure 9.** We illustrate the importance of the radial advection in our model showing the fraction of the energy liberated in the disc as a function of accretion rate without advection (continuous lines) and with advection term included (dashed lines); the corona dissipates a fraction $1 - \xi$. Curves are labelled by the viscosity parameter $\alpha$. For better clarity, only two cases with advection are shown ($\alpha = 0.1$ and 1.0). Other parameters: $M = 10^8 \, {\rm M}_\odot$, $r = 10 \, R_{\rm Schw}$.

Therefore the only result of the advection term is to extend the solutions slightly towards higher accretion rates and to lower the fraction of the energy liberated in the disc for a given accretion rate, but the differences are practically negligible. To allow the comparison with the previous case we show the efficiency plot for two values of the viscosity parameter $\alpha$ (Fig. 9). Other properties of the corona also do not practically differ from those computed neglecting the advection term (see Fig. 4) so we do not repeat the plots in this Section.

Our results strongly differ from advection-dominated solutions for the entire disc. This is due to the fact that stratified flow is cooled far more efficiently than an unstratified one as the optically thick layer supplies copious photons to the corona.

## 5 DISCUSSION AND CONCLUSIONS

### 5.1 Luminosity states

The dependence of the disc/corona solution on accretion rate at a given radius show three distinct states. At very low accretion rates the corona does not develop, at intermediate accretion rates a strong corona forms and at high accretion rates the corona becomes relatively weak. Such states may correspond to observed states: very low state, low state and high state.

Very low state was observed in the case of a galactic source V404 Cyg and suggested in the case of accretion at the central black hole of the Galaxy (Falcke & Heinrich 1994). Such a state is characterized by very soft thermal emission, possibly corresponding to standard accretion disc although the typical temperature is somewhat higher than expected on the basis of accretion rate appropriate for the observed luminosity.

Low state is well observed in a number of black hole candidate galactic sources and show general similarity with the spectra of most Seyfert galaxies (see Mushotzky et al. 1993). Its characteristic property is the presence of hard X-ray power law with the energy index $\sim 0.9$ (or sometimes flatter) extending to a few hundreds of keV. This component contains considerable or even dominating fraction of the total bolometric luminosity of the source.

High states are also well defined in the case of black hole galactic candidates. The spectrum is again relatively soft, most probably due to the thermal emission of optically thick medium and it is approximately modeled by accretion disc although the temperatures are somewhat higher than predicted by standard accretion discs due to the Comptonization inside optically thick discs. Similar spectra are characteristic for quasars and exceptional Seyfert galaxies (e.g. RE 1034+39, Puchnarewicz et al. 1995; Pounds, Done and Osborne 1995).

The coincidence of these trends with the predictions of our model is encouraging although several simplifications were made in our computations so it is difficult to expect detail quantitative agreement between the data and the model a this stage.

### 5.2 Spectra

The optical/UV/X-ray spectra computed on the basis of our model are in agreement with the trends observed in the data. The optical spectra are softer than predicted by standard accretion disc model due to the significant mass loss from the disc surface and subsequent decrease of the accretion rate towards the central black hole. The slope of the hard X-ray power low is also reproduced for external values of accretion rate about 0.01 of the Eddington value due to formation of the strong corona.

However, our results should not be treated as a proof that the accreting continuous corona is actually the best explanation of the origin of the X-ray emission. Although the model predictions at a given radius do not crucially depend on the assumptions, including the value of the viscosity parameter, the predictions of the radial dependencies, particularly of the mass loss, do depend strongly on the viscosity. Further work is needed before any distinction between accreting continuous and patchy magnetically driven corona (e.g. Haardt et al. 1994, Stern et al. 1995) could be made.

### 5.3 Outflow

In our model the accretion flow into the black hole is accompanied by very strong outflow. Such a significant outflow of the material seems to be one of the observed basic characteristics of nuclear activity. Radio-loud objects have collimated outflows whilst similar but uncollimated outflow is argued for in radio-quiet objects (e.g. Stocke et al. 1992), both quasars and Seyfert galaxies.

There are several direct evidences of the outflow from the central regions of active galaxies although the localization of the origin of the outflow is hardly constrained. Absorption features due to the partially ionized medium present along the line of sight to the nucleus (i.e. warm absorber) detected in a number of Seyfert 1 and Narrow Line Seyfert galaxies (e.g. Halpern 1984; Nandra & Pounds 1994; Fiore et al. 1993; Turner et al. 1993) are most probably due to the outflowing gas (Reynolds & Fabian 1995). Similar features are observed in BL Lac objects (Madejski et al. 1991). ASCA observations of NGC 4051 (Mihara et al. 1994) suggest the velocity $\sim 10000$ km/s and lower values ($\sim 1000$ km/s) were determined for NGC 4151 (Kriss et al. 1992). Evidences of an outflow with velocities $\sim 3500 - 5000$ km/s are also seen in Seyferts PG 1351+64, Ton 951 and III Zw 2 (Stocke et al. 1994). The significant change of absorption features in three weeks observed for MCG-6-30-15 (Fabian et al. 1994) indicate that the warm absorber is located within a few light weeks. It may have the covering factor close to unity (e.g. George, Turner & Netzer 1995 for NGC 3783). Further out the outflow may continue through the BLR (although in the case of the wind from the inner edge of the dusty/molecular torus the outflow is not a continuation of the nuclear wind, Krolik & Begelman 1986, Balsara & Krolik 1995) to NLR (e.g. Krolik & Vrtilek 1984; see also Wilson 1994, Mulchaey, Tsvetanov & Wilson 1994). The mass flux in that region was estimated as $\sim 1 M_\odot/yr$ (Krolik & Vrtilek 1984).

Indirect support for the nuclear wind comes from the Broad Line Region model developed by Smith & Raine (1985) and Mobasher & Raine (1989). Clouds in their model forms due to the collision of the nuclear wind with another wind developed from outer parts of the disc. They expect nuclear wind velocities of order of $10000 - 50000$ km/s on the basis of line profiles and intensity ratios. The velocities of the outflow of the nuclear wind predicted by our model are also in that range which is encouraging for future studies.

### ACKNOWLEDGMENTS

This project was partially supported by PICS/CNRS no. 198 "Astronomie Pologne", by French Ministry of Research and Technology within RFR programme, by grant no. 2P30401004 of the Polish State Committee for Scientific Research and in part by a postdoctoral fellowship of the Deutsche Forschungsgemeinschaft (DFG) under Gz. Mu 1020/3-1.

## APPENDIX A

We show here how equations (9) and (10) can be partly solved analytically. Surprisingly the second differential equation (10) can be integrated without knowing the solution for $\dot{m}$ explicitly. We only need to multiply the second differential equation by an integrating factor which is $P^{-3}$ in this case.

We now obtain an implicit function in $P$ for the solution of the differential equation:

$$F(z,P) = C_1 + \frac{\dot{m}^2 kT}{2P^2 m_{\rm H}} - \frac{F_{\rm soft}\kappa_{\rm es} m_{\rm H}}{ckT}z \\ - \frac{R_{\rm Schw}c^2 m_{\rm H}}{2kT}\frac{1}{\sqrt{r^2+z^2}} + \ln P = 0 \quad (A1)$$

where $C_1$ is the integrating constant. If the solution shall go through the sonic point we obtain for this case

$$F(z,P) = -\frac{1}{2} + \frac{\dot{m}^2 kT}{2P^2 m_{\rm H}} - \frac{F_{\rm soft}\kappa_{\rm es}m_{\rm H}}{ckT}(z - z_{\rm sonic}) - \\ \frac{R_{\rm Schw}c^2 m_{\rm H}}{2kT}\left[\frac{1}{\sqrt{r^2+z^2}} - \frac{1}{\sqrt{r^2+z_{\rm sonic}^2}}\right] + \\ \ln\left(\frac{P}{P_{\rm sonic}}\right) = 0. \quad (A2)$$

To obtain a solution for $\dot{m}$ and subsequently for $P$ we assume first that the viscosity $\alpha = 0$, i.e. $v_r = 0$. In this case the two differential equations decouple and we can solve for $\dot{m}$. For the solution we obtain

$$\dot{m} = \frac{C_2}{r^2 + z^2}, \quad (A3)$$

where $C_2$ is the second integrating constant. To solve the differential equations for $\alpha \neq 0$ we take the ansatz

$$\dot{m} = \frac{f(z)}{r^2 + z^2}, \quad (A4)$$

which yields

$$\frac{df}{dz} = \frac{2v_r r m_{\rm H}}{kT}p(z) \quad \text{or} \\ f(z) = C_2 + \frac{2v_r r m_{\rm H}}{kT}\int_{z_0}^{z} P(z)dz. \quad (A5)$$

The first term in this equation represents the mass flux from the disc to the corona whilst the second term gives the net effect of the radial flow in the corona itself.

In principle we can insert the complete solution for $\dot{m}$ into equation (A1) which yields us an equation only for $P$. This procedure yields us an integral equation for $P$ which can be transformed into an implicit differential equation only in $P$.

Since the solutions for $P(z)$ for different viscosity parameters $\alpha$ do not differ very much from the solution for $\alpha = 0$ we can interpret $\alpha$ as a small distortion and try to obtain an approximate solution by Taylor expansion.

For low accretion rates ($\dot{M} < 0.1 \times 10^{26}$ g/s) we can neglect the term $\dot{m}^2 kT/2m_\mathrm{H} P^2$ in equation (A2) for small $z$ and obtain a very good approximation for the pressure term

$$P(z) = P_\mathrm{sonic} \exp\left[\frac{1}{2} + \frac{F_\mathrm{soft}\kappa_\mathrm{es}}{cT}(z - z_\mathrm{sonic}) + \frac{R_\mathrm{Schw}c^2 m_\mathrm{H}}{2kT}\left(\frac{1}{\sqrt{r^2+z^2}} - \frac{1}{\sqrt{r^2+z_\mathrm{sonic}^2}}\right)\right] \quad \text{(A6)}$$

for $z_0 \leq z \lesssim z_\mathrm{sonic}$. The error at the location $z_0$ is usually much smaller than 1%. The error increases with larger $z$ and at the location $z_\mathrm{sonic}$ in equation (A6) the pressure is overestimated by a factor $e^{1/2}$. However, for very low accretion rates ($\dot{M} < 0.01 \times 10^{26}$ g/s) equation (A6) yields excellent results because the relevant range for the pressure terms is in the range $z_0 \leq z \ll z_\mathrm{sonic}$, where $P$ is sufficient large.

In general we can state that the pressure decreases exponentially with larger $z$ in the range $z_0 \lesssim z \lesssim z_\mathrm{sonic}$. Considering equation (A4) and equation (A5), the same holds for $\dot{m}$ for $\alpha > 0$.

## APPENDIX B

In Appendix A we learned that we can decouple the two differential equations (9) and (10) for $\dot{m}(z)$ and $P(z)$ and obtain a differential equation only in $\dot{m}$ or $P$.

To compute a contour plot for $P(z)$ it is not at all clear that we can choose the second integrating constant $C_2$ in equation (A5) independently of the integrating constant $C_1$ in equation (A1).

We derive here the curve where $dP/dz \to \infty$ in the contour plots in Figure 1. This curve fully determines the initial values for $P$ and $\dot{m}$ of the contour plot if we determine the transonic solution. To obtain $dP/dz \to \infty$ in the contour plot we have the condition $P = \dot{m}\sqrt{kT/m_\mathrm{H}}$. Inserting now the solution of $\dot{m}$ of equations (A4) and (A5) yields

$$(r^2 + z^2)P/\sqrt{\frac{kT}{m_\mathrm{H}}} = C_2 + \frac{2v_r r}{kT/m_\mathrm{H}} \int_{z_0}^{z} P(z)dz \quad \text{(B1)}$$

After differentiating equation (B1) we obtain

$$\frac{dP}{dz}(r^2 + z^2) + 2zP = \frac{2v_r r}{\sqrt{kT/m_\mathrm{H}}}P. \quad \text{(B2)}$$

For this case the variables can be easily separated and the differential equation can be solved. Finally we obtain the solution,

$$P_\mathrm{initial}(z) = \frac{C_3}{r^2 + z^2} \exp\left[2v_r \left(\frac{kT}{m_\mathrm{H}}\right)^{-1/2} \arctan(\frac{z}{r})\right]. \quad \text{(B3)}$$

The integrating constant $C_3$ has to be calibrated in such a way that $P_\mathrm{initial}(z_\mathrm{sonic}) = P_\mathrm{sonic}$. The initial values for $\dot{m}$ can be determined subsequently by the condition $\dot{m} = P/\sqrt{kT/m_\mathrm{H}}$.

## APPENDIX C

We can derive simple analytical expressions for the parameters of the corona as a function of a mass of the central black hole, $M$, disc radius $x$ measured in Schwarzschild units ($x \equiv r/R_\mathrm{Schw}$) and accretion rate in units of the Eddington accretion rate with efficiency 1/16. For that purpose we assume that the dynamical terms and the advection term are negligible, the ion temperature is much larger than the electron temperature, the corona structure can be approximated using vertically averaged quantities and the nonlinear terms in energy balance equations are negligible. We also approximate $\ln((3-\xi)/(1+\xi))$ with $1-\xi$ which is accurate for $\xi \sim 1$). The boundary condition is included through $f(r)$ (see equation 14a).

Under these assumptions the parameters of the corona are:

$$1 - \xi = 0.017\alpha^{-1/4}f(r)^{-1/4}\dot{m}^{-1/4}x^{3/8}, \quad \text{(C1)}$$

$$T_\mathrm{i} = 5.2 \times 10^{12}\alpha^{-5/4}f(r)^{3/4}\dot{m}^{3/4}x^{-9/8}, \quad \text{(C2)}$$

$$T_\mathrm{e} = 1.3 \times 10^{8}\alpha^{-1/4}f(r)^{-1/4}\dot{m}^{-1/4}x^{3/8}, \quad \text{(C3)}$$

$$y = 1 - \xi \quad \text{(C4)}$$

and

$$\tau = 0.18. \quad \text{(C5)}$$

As we see, the properties of the corona do not depend on the mass of the central object so the same model can be applied to AGN and galactic black hole candidates. This is in contrast with the cold disc properties like temperature or the gas pressure to the total pressure ratio.

## APPENDIX D

In this Appendix we derive a set of equations which in first order represents the global solutions for the Figs. 3 and 4. The equations are needed to obtain the results for different $r$ with higher accuracy that provided by formulae given in Appendix (C). This higher accuracy is essential to calculate the vertical vs. radial flow (see equation 26) for different radii. Since equation (A6) represents a very good solution for the pressure for small $z$ we have $P(z_0) \approx P(0) = P_0$ and we can write for $z \ll z_\mathrm{sonic}$

$$\frac{1}{\sqrt{r^2+z^2}} \approx \frac{1}{r}(1 - \frac{z^2}{2r^2}),$$

which yields

$$P_\mathrm{approx}(z) = P_0 \exp\left[-\frac{MGm_\mathrm{H}}{kT_\mathrm{i}}\frac{z^2}{2r^3}\right]. \quad \text{(D1)}$$

In addition we neglected the term with $F_\mathrm{soft}$ in equation (D1). Equation (D1) is a good approximation for small $z$, but a bad one for large $z$. However, the overall integral over the pressure should represent a good approximation because the main quantity of the integral is contributed by small $z$. Therefore the decrease of the function for large $z$ is not important for this purpose. Using now the relation

$$\int_0^\infty P_{\text{approx}}(z)dz = \frac{P_0}{\Omega_K}\sqrt{\frac{\pi}{2}}\sqrt{\frac{kT_i}{m_H}} \tag{D2}$$

and the relation

$$\int_0^\infty P^2_{\text{approx}}(z)dz = \frac{P_0^2}{\Omega_K}\frac{\sqrt{\pi}}{2}\sqrt{\frac{kT_i}{m_H}} \tag{D3}$$

yields us for equation (16)

$$(1-\xi)F_{\text{tot}} = \frac{3}{2}\alpha P_0 \sqrt{\frac{\pi}{2}}\sqrt{\frac{kT_i}{m_H}} \tag{D4}$$

and for equations (20) and (21)

$$\ln\frac{3-\xi}{1+\xi} = \sqrt{\frac{\pi}{2}}\frac{P_0}{\Omega_K}\frac{\kappa_{\text{es}}}{\sqrt{kT_i/m_H}}\frac{4kT_e}{m_e c^2}\left(1 + \frac{4kT_e}{m_e c^2}\right). \tag{D5}$$

Further we obtain for equation (17)

$$(1-\xi)F_{\text{tot}} = \frac{3\sqrt{\pi}}{4}\frac{\sqrt{m_H}(T_i - T_e)}{\sqrt{k}T_e^{3/2}T_i^{3/2}}\frac{P_0^2}{\Omega_K}$$
$$\times 2.44 \times 10^{21} \ln\Lambda \left[1 + \left(\frac{4kT_e}{m_e c^2}\right)^{1/2}\right]. \tag{D6}$$

Finally we need a fourth equation, which is represented by equation (22) and which can be taken without alteration.

$$\Xi_0 \left(\frac{T_e}{10^8 \text{K}}\right)^{-3/2} = \frac{(1-\xi)F_{\text{tot}}}{2cP_0} \tag{D7}$$

Equations (D4)–(D7) represents the set of equations which yields us the global solution for the set of parameters $M$, $\dot{M}$, $\alpha$, $r$. The four equations can be solved easily numerically and they yield us the four unknown quantities $\xi$, $T_i$, $T_e$ and $P_0$. Also equations (D4)–(D7) can be used to derive the equations in Appendix C. For example, equations (D4) and (D7) yield immediately that $T_i \propto T_e^{-3}$ in first order.

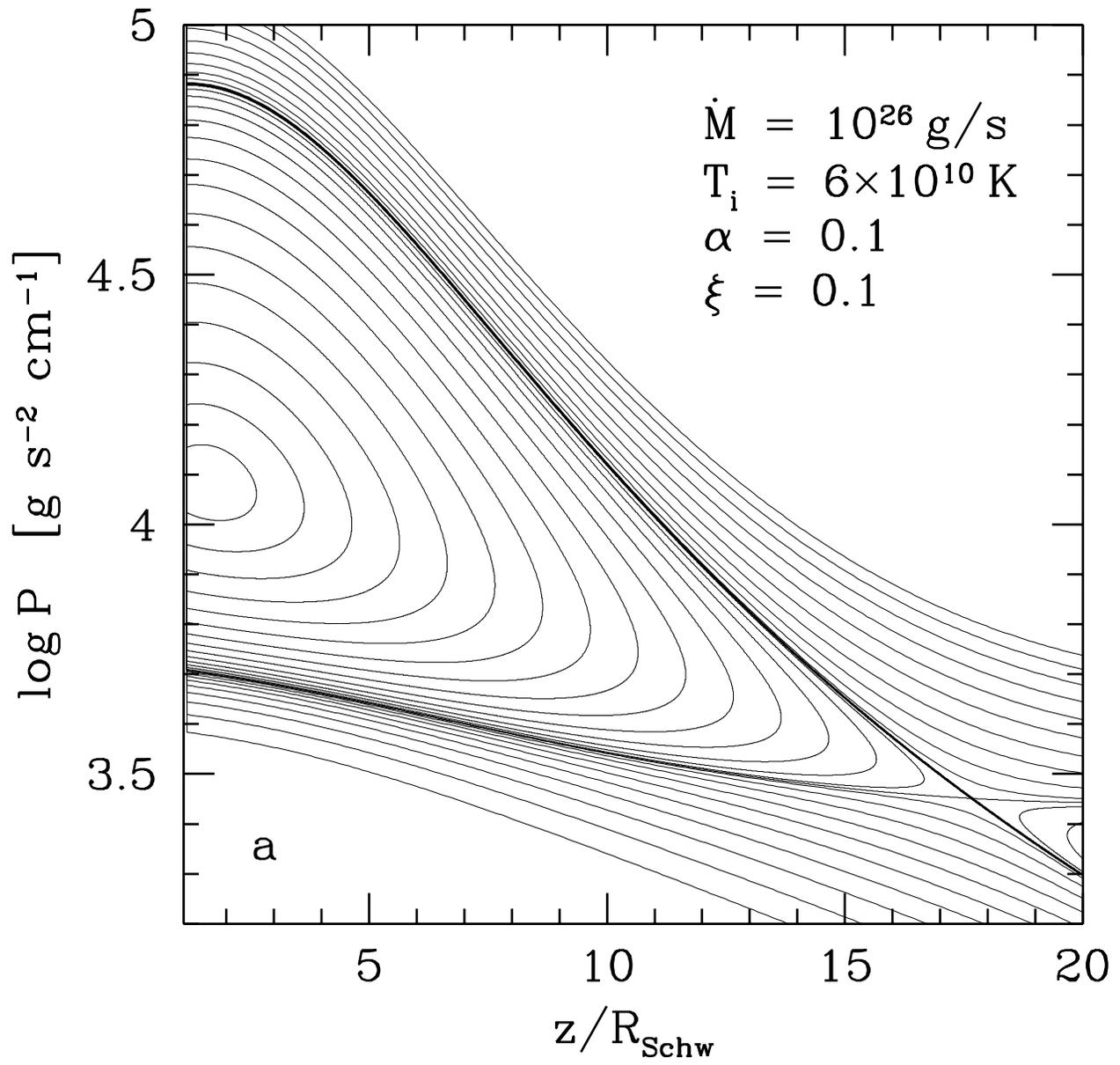

Fig. 1

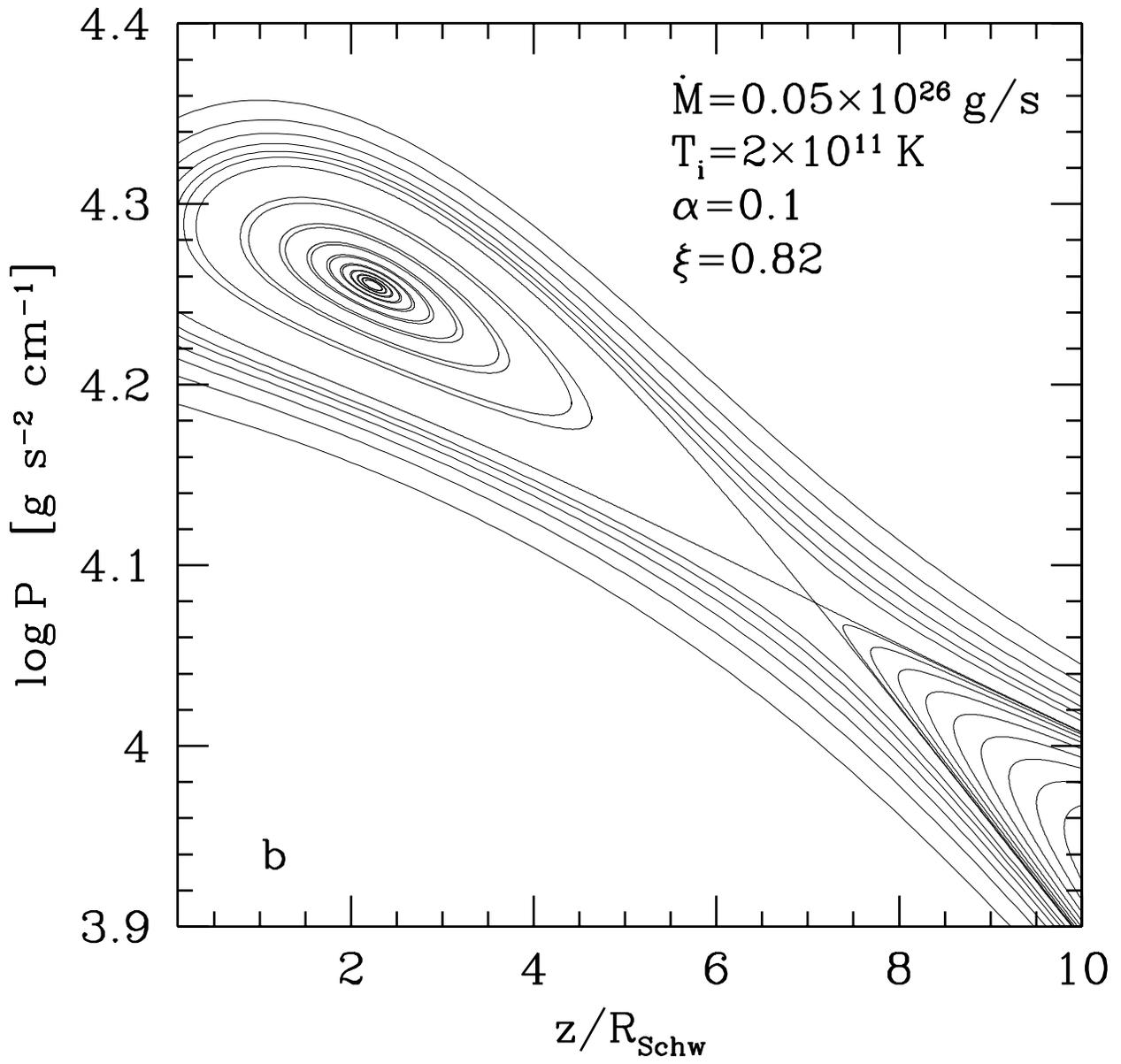

Fig. 1